\documentclass[prb,amssymb,preprint,floatfix]{revtex4-1} 

\usepackage{soul}
\usepackage{ulem}
\usepackage{xcolor}
\usepackage{amsmath}  % needed for \tfrac, \bmatrix, etc.
\usepackage{amsfonts} % needed for bold Greek, Fraktur, and blackboard bold
\usepackage{graphicx} % needed for figures

\begin{document}
    
    %%%%%%%%%%%%%%%%%%%%%%%%%%%%%%%%%%%%%%%%%%%%%%%%%%%%%%%%%%%%%%%%%%%%%
    %                        Title   
    \title{Black-hole analog in vehicular traffic}
    
    %%%%%%%%%%%%%%%%%%%%%%%%%%%%%%%%%%%%%%%%%%%%%%%%%%%%%%%%%%%%%%%%%%%%%
    %                 Authors & Addresses 
    
    \author{Luanna K. de Souza}\email{luanna.souza@unesp.br}    
    \affiliation{Instituto de F\'\i sica Te\'orica, 
    Universidade Estadual Paulista, Rua Dr.\ Bento Teobaldo Ferraz, 271, 01140-070, S\~ao Paulo, S\~ao Paulo, Brazil}
    
    \author{George E. A. Matsas}\email{george.matsas@unesp.br}
    \affiliation{Instituto de F\'\i sica Te\'orica, 
    Universidade Estadual Paulista, Rua Dr.\ Bento Teobaldo Ferraz, 271, 01140-070, S\~ao Paulo, S\~ao Paulo, Brazil}
        
    %%%%%%%%%%%%%%%%%%%%%%%%%%%%%%%%%%%%%%%%%%%%%%%%%%%%%%%%%%%%%%%%%%%%%
    %                    Abstract       
    
    \begin{abstract} 
   We propose here a simple black-hole analog in vehicular-traffic dynamics. The corresponding causal diagram is determined by the propagation of the tail light flashes emitted by a convoy of cars on a highway. In addition to being a new black-hole analog,  this illustrates how causal diagrams, so common in general relativity, may be useful in areas as unexpected as vehicular-traffic dynamics.
    \end{abstract}
    
    %%%%%%%%%%%%%%%%%%%%%%%%%%%%%%%%%%%%%%%%%%%%%%%%%%%%%%%%%%%%%%%%%%%%%
    %                   Main text
    \maketitle
    
    %%%%%%%%%%%%%%%%%%%%%%%%%%%%%%%%%%%%%%%%%%%%%%%%%%%%%%%%%%%%%%%%%%%%%
    
    %%%%%%%%%%%%%%%%%%%%%%%%%%%%%%%%%%%%%%%%%%%%%%%%%%%%%%%%%%%%%%%%%%%%%
     \section{Introduction}
     \label{introduction}
    %%%%%%%%%%%%%%%%%%%%%%%%%%%%%%%%%%%%%%%%%%%%%%%%%%%%%%%%%%%%%%%%%%%%%

%``Everything is information"; this was John Wheeler's ultimate insight after his long and successful career. Although Wheeler's maxim was inspired by quantum mechanics, his motto is valid in all science. The scientific method only tests whether a theory can predict {\it outputs} from {\it inputs}; any underlying interpretation goes beyond it.  As a result, the issue of how information propagates is far-reaching from general relativity to vehicular-traffic dynamics. 

The area of analog models of gravitational systems has a splendid history. Its genesis can be traced back to the beginning of the 1980s with two independent papers: one by Vincent Moncrief~\cite{M80} and the other by Bill Unruh~\cite{U81}. Moncrief considered a perturbed fluid in a general relativistic spacetime and concluded that {\it one can study the stability of spherical accretion onto nonrotating stars by the methods used for black holes}, while Unruh~\cite{U81} showed that the perturbation of usual nonrelativistic irrotational fluids develops sonic horizons which emit sound waves with a thermal spectrum analogous to the radiation derived by Hawking for usual black holes~\cite{H74}. After that, various analog black holes were discovered in Bose-Einstein condensates, shallow water waves, slow light, and so on. (See Refs.~\onlinecite{V98, BLV11} for reviews of analog models of gravity.) Our vehicular-traffic black hole adds a new and appealingly simple model to this list.

We propose a simple black-hole analog in vehicular-traffic dynamics. The corresponding causal diagram is determined by the propagation of the tail light flashes emitted by a convoy of cars on a straight highway. Similar to a real black hole, our vehicular-traffic black-hole model exhibits event and apparent horizons and a Planck region where the causal structure breaks down. This model also illustrates how (informational) causal diagrams, so common in general relativity, may be useful in areas as unexpected as vehicular-traffic dynamics. It is possible that this simple physical-informational approach, complemented by more traditional ones~\cite{ID16}, could help to avoid traffic accidents under impaired visibility conditions. If this occurs, it would constitute a quite curious relativity spin-off.

The paper is organized as follows. In Sec.~\ref{model} we present our vehicular-traffic model. In Sec.~\ref{stationary regime} we present the information-flow metric associated with the propagation of the brake-light signal in the simplified time-independent (stationary) regime. In Sec.~\ref{vehicular-traffic black hole} we show how the drivers' reaction time can give rise to a vehicular-traffic black hole. In Sec.~\ref{singularity} we show that taking into account accidents on the track drives us to the nonstationary regime. In Sec.~\ref{nonstationary regime} we work out the more realistic nonstationary regime. In Sec.~\ref{results} we summarize our results and add some final remarks. 
    
    %%%%%%%%%%%%%%%%%%%%%%%%%%%%%%%%%%%%%%%%%%%%%%%%%%%%%%%%%%%%%%%%%%%%%
     \section{The vehicular-traffic model}
     \label{model}
\begin{figure}[t]
       \centering
       \includegraphics[width=90mm]{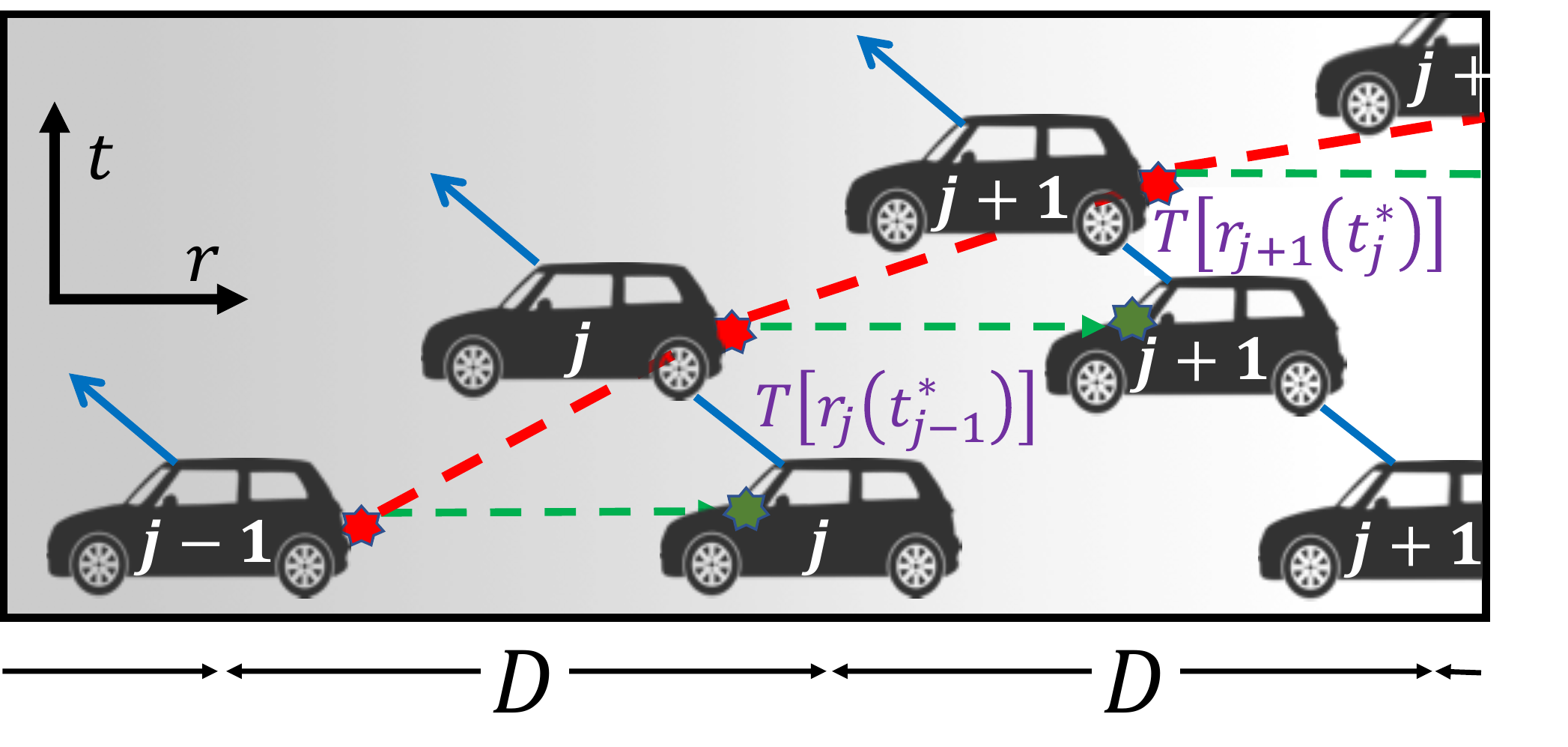}
        \caption{(Color online) Convoy of vehicles, each separated by a distance~$D$, each moving with the same constant velocity $v$, on a one-dimensional road. The stars at the rear-lamp denote the first-flash due to braking which occurs at the spacetime coordinates $(t^*_j,r^*_j)$; the stars at the windshield denote the instantaneous arrival of the light ray at the next vehicle; and the lines joining the rear lamps describe the propagation of the braking information. Once the rear lamp of the $j$-th vehicle is applied, the next one, $j+1$-th, will take a time-lapse $T(r_{j+1})$ to react. The more impaired the visibility conditions, the longer the reaction time. It is assumed that such information propagates as a chain from one vehicle to the next on the right. (For the sake of later comparison with our black-hole spacetime diagram, the convoy is shown to move in the negative $r$~direction.)}
       \label{carros2}
\end{figure}

    %%%%%%%%%%%%%%%%%%%%%%%%%%%%%%%%%%%%%%%%%%%%%%%%%%%%%%%%%%%%%%%%%%%%%
Consider a caravan of equally spaced inertial cars initially moving (to the left) with velocity $v_j={\rm const},\; j = 1, 2, \ldots $ on a straight highway covered with Cartesian coordinates. The vehicles are assumed to be smaller than any other distance scale in the problem. We also assume that the brake lights can only be seen by adjacent vehicles and that, once the car ahead brakes, the car behind should also brake after some reaction time $T>0$. $T$ will vary depending on the weather conditions. The more fog is present, the larger~$T$ will be. We are interested in following the propagation of the tail light flashes along the highway when the brakes are applied in succession (see Fig.~\ref{carros2}). 
     
We shall denote by $r_j(t)$ the position of the $j$-th car at the instant~$t$. The propagation of the tail light flashes can be represented by the sequence
     \begin{equation}
     \{(t^*_1,r^*_1),\ldots,(t^*_j,r^*_j),\ldots \}, 
     \label{worldline}
     \end{equation}
where $(t^*_j,r^*_j)$ denotes the instant $t^*_j$ and position $r^*_j$ when the $j$-th vehicle starts to brake. 
   
   %%%%%%%%%%%%%%%%%%%%%%%%%%%%%%%%%%%%%%%%%%%%%%%%%%%%%%%%%%%%%%%%%%%%%
   \section{The stationary regime}
   \label{stationary regime}
   %%%%%%%%%%%%%%%%%%%%%%%%%%%%%%%%%%%%%%%%%%%%%%%%%%%%%%%%%%%%%%%%%%%%%
   
  Let us begin by assuming the braking is smooth and brief enough such that $v_j =v \approx {\rm const}$. Hence, the initial separation between contiguous vehicles $ | r_{j+1}(t) - r_j(t) | \equiv D $ remains (approximately) the same even after the brakes are applied, avoiding shocks. In addition, we start with the assumption that $T=T(r)$ is  simply a monotonic function of the position where the driver receives the tail light signal from the car ahead. Later on, we will remove this restriction and envisage the more realistic case where the visibility conditions vary also in time: $T=T(t,r)$.
     
 We can think of the caravan as a sort of medium propagating with velocity~$v$ on a preferred reference frame~${\cal S}$, namely, the highway. 
 %This is a convenient (but not necessary) simplification that will smooth the link of the straight dashed segments shown in Fig.~\ref{carros2}. This is justified under the present philosophy of obtaining a global, rather than local, understanding of the information flow. 
 The speed of the information propagation with respect to the vehicular medium is
    \begin{equation}
    c_s = D/T(r) > 0, 
    \label{rep}
    \end{equation}
where $T\sim 1~{\rm s}$ under normal visibility conditions.\cite{JR71,H01}  
Now, let us note that, since all the velocity scales in the  problem are much smaller than the speed of light, we can assume Newtonian physics all around in our calculations. Thus, we use the usual nonrelativistic velocity composition to express the information-propagation velocity with respect to the highway ${\cal S}$:
    \begin{equation}
        dr/dt = \mp c_s + v, \quad v\gtrless 0,  
        \label{star1}
    \end{equation}
where we will choose here $+c_s$ and $v<0$ in Eq.~\eqref{metric1} in accordance with Fig.~\ref{carros2}. Note that we are only interested in the ``outgoing" information flow, {\it i.e.}, from the $j$-th to the $j+1$-th car. 

Now, we aim to describe the information flow as propagating in a vehicular-traffic effective spacetime. To define any physical space or spacetime, we must know its metric. Metrics allow us to measure distances between neighboring points. For instance, the so-called ``line element'' associated with the metric of the usual Galilean space (of Newtonian mechanics) is
\begin{equation}
  dl^2 = dx^2 + dy^2 + dz^2,  
  \label{Galileo}
\end{equation}
where $\{ x, y, z\}$ are usual Cartesian coordinates, since the squared spatial distance between any pair of neighboring points is $dl^2$. In the relativistic realm, neither space nor time have an absolute meaning on its own right. Rather, only a combination of both of them does. Which combination depends on the case. Far from matter, singularities and other oddities, where inertial observers experience a homogeneous and isotropic space, the line element is the Minkowski one (of special relativity):
\begin{equation}
ds_{M}^2 = -c^2 dt^2 + dx^2 + dy^2 + dz^2, 
\label{Minkowski}
\end{equation}
where $\{ t, x, y, z\}$ are spacetime Cartesian coordinates. It is important to notice that $ds_{M}^2$ is not positive definite. In particular, light rays follow ``null worldlines'' where neighboring points satisfy $ds_{M}^2=0$. 
  
Now, let us seek the spacetime where the information propagation~\eqref{star1} follows null worldlines. 
The line element of such a spacetime can be found by inspection: 
    \begin{equation}
    ds_{VT}^2 \equiv - (c_s^2 - v^2) \ dt^2 - 2v dt dr + dr^2. \label{metric1}
    \end{equation}
Indeed, Eq.~\eqref{star1} can be recovered from Eq.~\eqref{metric1} by imposing $ds_{VT}^2=0$. Thus, the propagation of the brake lights will be seen as null worldlines in the vehicular-traffic effective spacetime defined by the line element~\eqref{metric1}.
Note that for $v=0$, Eq.~\eqref{metric1} simply reads $ds_{VT}^2 = -c_s^2 dt^2+dr^2$ corresponding to the Minkowski line element (with $c_s$ playing the role of the speed of light) but for $|v| \neq 0$ nontrivial effects may come out because the right-moving information is dragged by the ``left-moving vehicular-traffic spacetime.'' 

Let us examine the situation shown in Fig.~\ref{carros2} and assume the visibility conditions deteriorate to the west (decreasing $r$ coordinates). Thus, the smaller the~$r$ coordinate the larger the reaction time $T(r)$. Now, let us consider that, as we move from far east to the west, there exists a point, $r=r_H$, where $T(r)$ is sufficiently large such that 
\begin{equation}
D/T(r_H) = |v|.
\label{apparent horizon}
\end{equation}
The coordinate $r=r_H$ fixes the location of an effective {\it apparent horizon} like that of a black hole: for $r<r_H$ originally outgoing information lines will propagate ``inwards'' with respect to the highway (although they will always propagate ``outwards'' with respect to the convoy). This is so because the information propagates so slowly in comparison to the cars motion that, by the time that car~$j+1$ reacts to the signal, it has already passed the point where the car~$j$ emitted the signal. This will preclude the tail light brake signals from ``properly'' propagating through the convoy, thus causing accidents. 

In the stationary regime, where $T=T(r)$ ($T$~does not depend on the coordinate~$t$), the apparent horizon will coincide with the {\it event horizon}, which is the black-hole frontier from inside which no information escapes. In general relativity, light rays emitted outwards at the event horizon of a black hole get frozen there.\cite{W84} 

At this point, one should notice that the existence of a vehicular-traffic black hole will only depend on a rapid enough deterioration of the visibility conditions as $r$~decreases (see Fig.~\ref{carros2}). In this case, the function~$D/T(r)$ will decrease rapidly as $r$ decreases, leading  $D/T(r) \to v $ at some $r=r_H$. Ideally, a realistic~$T(r)$ function would demand that data be experimentally collected on the ground but this is much beyond the scope of this work. Nevertheless, rather than choosing an arbitrary reaction time $T(r)$ to create a vehicular-traffic horizon and explore its nontrivial consequences, we show in Sec.~\ref{vehicular-traffic black hole} how $T(r)$ can be chosen to formally mimic the simplest black hole in nature: the Schwarzschild black hole, which is completely characterized by its mass~$M$. Readers who are not interested in this connection should just employ the $T(r)$ defined in Eqs.~\eqref{delay_E-F}, refer to Fig.~\ref{GRAF-3}, and skip directly to Sec.~\ref{singularity}.
    
    %%%%%%%%%%%%%%%%%%%%%%%%%%%%%%%%%%%%%%%%%%%%%%%%%%%%%%%%%%%%%%%%%%%%%
     \section{The vehicular-traffic black hole}
     \label{vehicular-traffic black hole}
    %%%%%%%%%%%%%%%%%%%%%%%%%%%%%%%%%%%%%%%%%%%%%%%%%%%%%%%%%%%%%%%%%%%%%
   
Let us now relate Eq.~\eqref{metric1} to the line element of a (Schwarzschild) static vacuum black hole with mass~$M$. The line element of a Schwarzschild black hole (omitted the angular part) can be cast in advanced Eddington-Finkelstein (E-F) coordinates~$(\bar{t},r)$ as~\cite{E24,F58}
        \begin{equation}
         ds_{BH}^2 =  -  C^2 f_-\, d\bar{t}^2  
                 + 2 C \, (r_H / r) \, d\bar{t} dr  
                 +  f_+\,  dr^2,
                 \quad \quad
               f_\pm (r) = 1 \pm r_H /r,
         \label{E-F-1}
        \end{equation}
where
$$
  r_H\equiv 2GM/C^2
$$ 
is the event-horizon radius, and we replaced the speed of light~$c$ by a convenient analogue speed  $C=200~{\rm km/h}$.  The similarity between Eqs.~\eqref{metric1} and~\eqref{E-F-1} is unambiguous, but there is a relevant difference: while the latter is already all determined, the former depends on $c_s(r)$, which varies according to the visibility conditions. Here, we will select $T(r)$ that yields a ``Schwarzschild'' analog vehicular-traffic  black hole.  
     
     The outgoing radial null  worldlines, $ds^2=0$, for Schwarzschild black holes in E-F coordinates satisfy 
     \begin{equation}
        dr/ d\bar{t} = C (r - r_H)/(r + r_H). 
        \label{dr_dt-E-F}
     \end{equation}    
     Integrating this, we obtain the corresponding wordlines, 
      \begin{equation}
       C \bar{t}  = r + 2r_H \ln \left| \frac{r}{r_H} -1 \right| + {\rm const}.
        \label{Photon}
     \end{equation}  
Now, comparing Eqs.~\eqref{star1} and~\eqref{dr_dt-E-F}, we choose $T(r)$ such that the right-hand sides of both of these equations are formally equal for 
$$
r>r_P \equiv r_H \, ( 1 - |v|/C)/( 1 + |v|/C).
$$
Then,
    \begin{equation}
        T(r) = (D/C) \left[ (r - r_H)/(r + r_H) + {| v |}/{C} \right]^{-1},\; r>r_P, 
        \label{delay_E-F}
    \end{equation}
and the reaction time~$T(r)$ is assumed to be arbitrarily large in the interval $ 0\leq r\leq r_P$. Note that $T(r)\to \infty$ as $r \rightarrow r_H$. It is important to keep in mind that our information-flow analysis will be restricted to the domain $r\geq0$, although the highway is assumed to be unbounded, $r\in (-\infty,+\infty)$.
\begin{figure}[t]
       \centering
       \includegraphics[width=60mm]{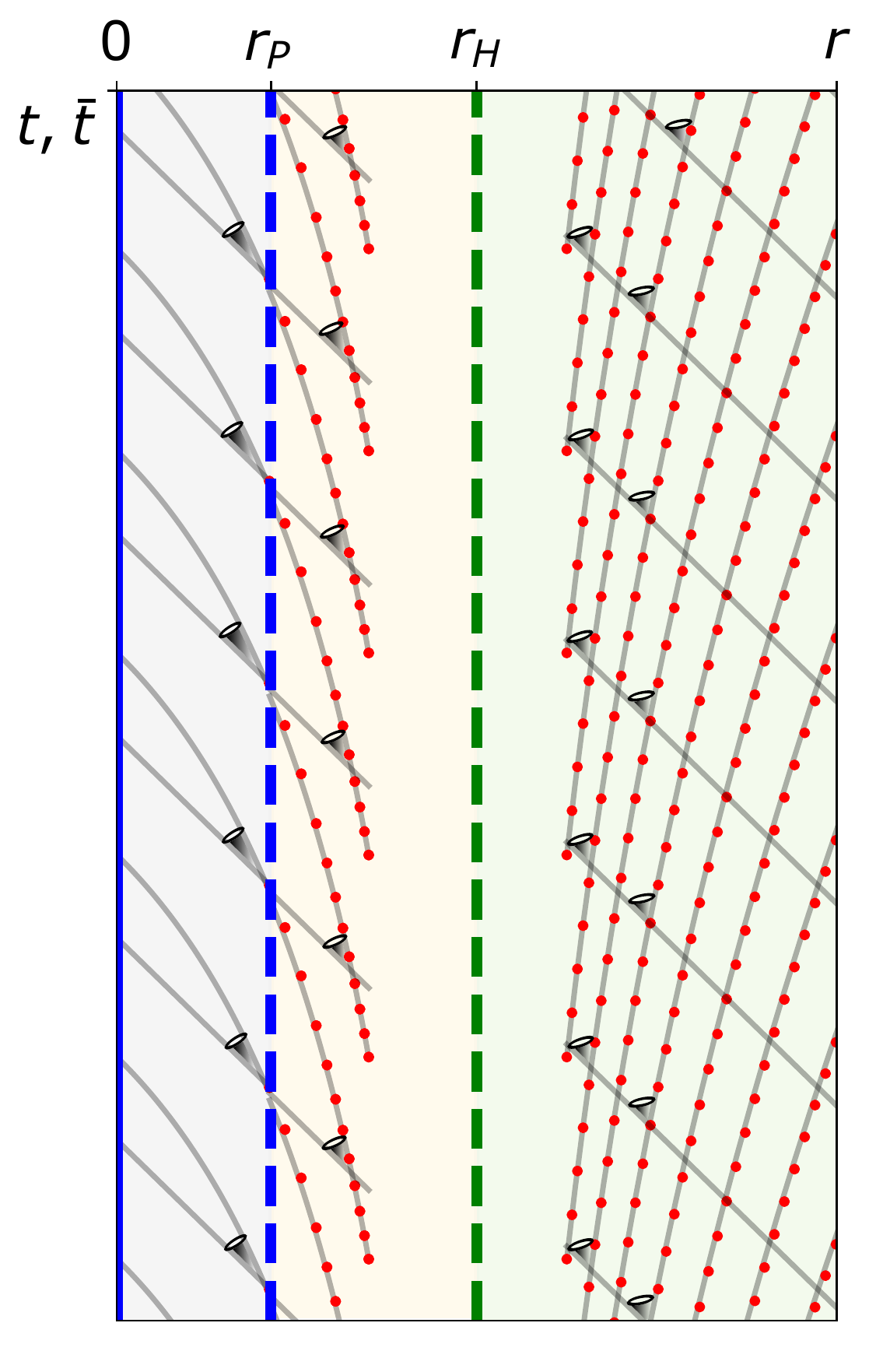}
        \caption{(Color online) The solid gray lines represent ingoing and outgoing null worldlines associated with the light cones of the Schwarzschild black hole. The dotted curves represent the propagation of the outgoing brake-light signals. Each dot represents the brake lights of 10 cars. Distinct lines are associated with different first-braking events: $(t_1^*,r_1^*)$. The outgoing black-hole and brake-light lines are nearly aligned. The vertical dashed line at $r= r_H$ marks the black-hole frontier. The outgoing information lines propagate inwards for $r_P<r<r_H$. The area $0\leq r\leq r_P$ plays the role of an effective information Planck region where $T(r)$ diverges and the causal structure collapses. Here,  $D= 0.1\, {\rm km}$, $v= 80\, {\rm km/h}$, and $r_H= 20\, {\rm km}$.}
       \label{GRAF-3}
    \end{figure} 
    
Given the first-car brake application, $(t_1^*, r_1^*)$, Eq.~\eqref{delay_E-F} determines the complete worldline~\eqref{worldline} through 
    \begin{equation}
    t^*_{j}  =  t^*_{j-1} + T [r_{j}(t^*_{j-1})], \quad
    r^*_j =  r_j(t_1^*) - | v | \, (t_j^* - t_1^*),  
    \label{Code}
    \end{equation}
where $r_j(t_1^*) = r_1^* + (j-1)D$. 

In Fig.~\ref{GRAF-3} we superimpose the smooth gray lines of the black-hole radial outgoing null worldlines in~$(\bar{t}, r)$~coordinates~\eqref{Photon} (with ${\rm const} = C t_1^* - r_1^* - 2 r_H \ln \left| r_1^*/r_H -1 \right|$) on the dotted lines of our brake-light lines in the usual~$(t,r)$ Cartesian coordinates~\eqref{Code}. The reason why the {\it continuous but nonsmooth} dotted lines of the brake lights overlap so well with the {\it smooth} gray lines of the black-hole outgoing worldlines is because we are dwelling in the small-$D$ regime: $D/c_s \cdot |dc_s/dr| \ll 1$. (This is in agreement with our assumption that the caravan should be seen as a sort of ``vehicular medium''.)  The vertical broken line at $r=r_H$ corresponds to the black-hole horizon, which means that the visibility conditions at $r_P<r<r_H$ slow down drivers' reactions enough to make outgoing information lines propagate ``inwards'' with respect to the highway. The gray region, $0 \leq r \leq r_P$, was excised from our information-flow graph since $T(r)$ diverges at $r = r_P$: vehicles that get the front-car brake signal at $0 \leq r \leq r_P$ do not react ever. In general, the brake-light line will terminate somewhere in $r < r_P $ due to cars that receive the front-car tail light flash at $r\gtrsim r_P$, braking at $r<r_P$, eventually. In analogy to relativity, the region $0\leq r \leq r_P$ will play the role of an information Planck region, where the theory will be assumed to break down. It is commonly assumed that black hole singularities are mathematical artifacts stemming from the fact that general relativity does not comply with quantum rules. Whether or not this is the case, we do not know. Nevertheless, it is fair to say that we shall not trust the causal structure predicted by general relativity close enough to such extreme regions as given by the Planck scale~$\sqrt {G \hbar/ c^3} \sim 10^{-33}~{\rm cm}$.~\cite{A10}
     
We must emphasize that despite all the similarities between Eqs.~\eqref{metric1} and~\eqref{E-F-1}, they only concern the propagation of information. The vehicle worldlines are not bound to any light cones shown in Fig.~\ref{GRAF-3}. In the stationary regime in which we are dwelling so far, $v_j=v={\rm const}$ and nothing would stop the vehicles from crossing the $r=0$ axis. However, the appearance of singularities due to collisions begs us to go beyond the stationary regime, changing this state of affairs.
      \begin{figure}[t]
       \centering
       \includegraphics[width=52mm]{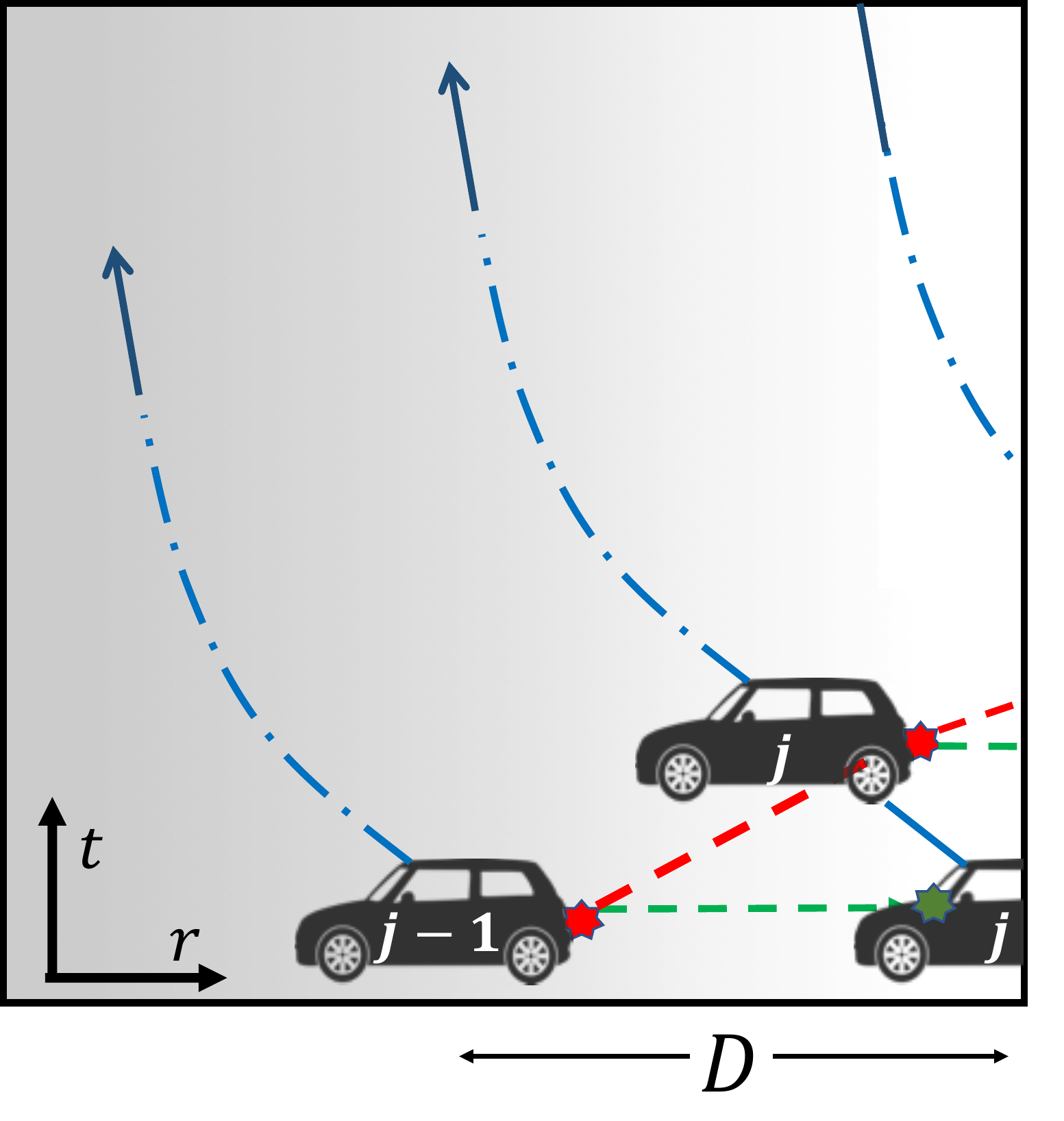}
        \caption{(Color online) Here vehicles~$j$ and~$j+1$ brake and decelerate, as shown by the blue dot-dashed lines. Eventually, they reestablish uniform motion without collision. The red dashed line represents the information flow world line and the green horizontal dashed line represents the instantaneous propagation of the brake light signal between adjacent cars. (Recall that all the action takes place in the usual Galileo spacetime.)}
       \label{semcolisao}
    \end{figure}
        \begin{figure}[ht]
       \centering
       \includegraphics[width=52mm]{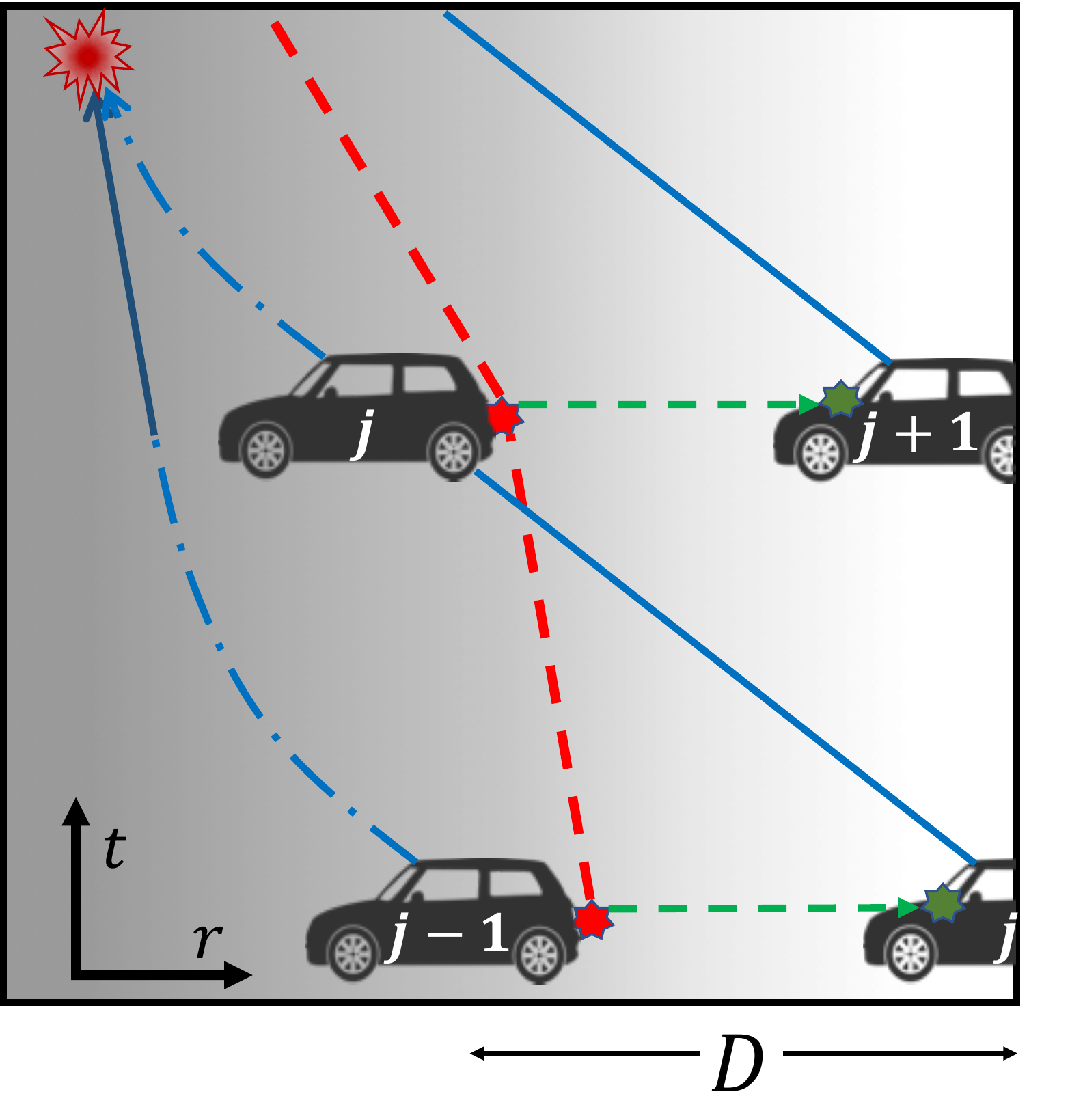}
        \caption{ (Color online) Vehicles~$j$ and~$j+1$ brake, decelerate, and eventually collide (stary red blob), forming a vehicular-traffic singularity. The fact that information associated with outgoing worldlines flows inwards (see red dashed segments) testifies that the situation occurs inside the vehicular-traffic black hole. The green horizontal dashed line represents the instantaneous propagation of the brake light signal between adjacent cars. (Recall that all the action takes place in the usual Galileo spacetime.)}
       \label{comcolisao}
    \end{figure}  
    %%%%%%%%%%%%%%%%%%%%%%%%%%%%%%%%%%%%%%%%%%%%%%%%%%%%%%%%%%%%%%%%%%%%%
     \section{Vehicular-traffic singularity}
     \label{singularity}
    %%%%%%%%%%%%%%%%%%%%%%%%%%%%%%%%%%%%%%%%%%%%%%%%%%%%%%%%%%%%%%%%%%%%%
   
Let us now assume that the application of the brakes decelerates the vehicle at a constant rate $a$, during the time interval
     \begin{equation}
         \Delta t =  \frac{|v| - |v_r|}{a},
         \label{Delta_t}
     \end{equation}
until it reaches some reduced safe speed~$|v_r|$, after which it proceeds with $v_r = {\rm const}$. (Recall that in our case $v, v_r<0$ and so $a>0$.) Figure~\ref{semcolisao} depicts a situation where both cars, $j$ and~$j+1$, apply the brakes, attain a reduced velocity~$v_r$, and move on without colliding with each other. A direct comparison with Fig.~\ref{GRAF-3} indicates this happens {\it outside the hole}, $r>r_H$, since outgoing information flows outwards. In contrast, the snapshot in Fig.~\ref{comcolisao} happens {\it inside the hole} at $r_P < r <r_H$, since outgoing information flows inwards. Depending on the physical situation, the cars still may not collide, but generally they will. There are different possibilities the $j+1$-th car behind hits the $j$-th car ahead depending on whether or not the front car secures~$v_r$ and the back car starts braking. In~Fig.~\ref{comcolisao} the $j+1$-th car reacts fast enough to begin braking but not to avoid a collision with the $j$-th car, which attained the speed~$v_r$. 
     
The first collision of a vehicular pileup will be interpreted as the appearance of a singularity since the rest of the convoy will collide in succession at the same spot. The fact that the singularity forms at some given time, rather than having existed forever, demands we abandon our initial stationarity assumption. Hence, let us move on and consider the physical situation where $T=T(t,r)$.

    %%%%%%%%%%%%%%%%%%%%%%%%%%%%%%%%%%%%%%%%%%%%%%%%%%%%%%%%%%%%%%%%%%%%%
     \section{Nonstationary regime}
     \label{nonstationary regime}
    %%%%%%%%%%%%%%%%%%%%%%%%%%%%%%%%%%%%%%%%%%%%%%%%%%%%%%%%%%%%%%%%%%%%%

     \begin{figure}[t]
       \centering
       \includegraphics[width=61mm]{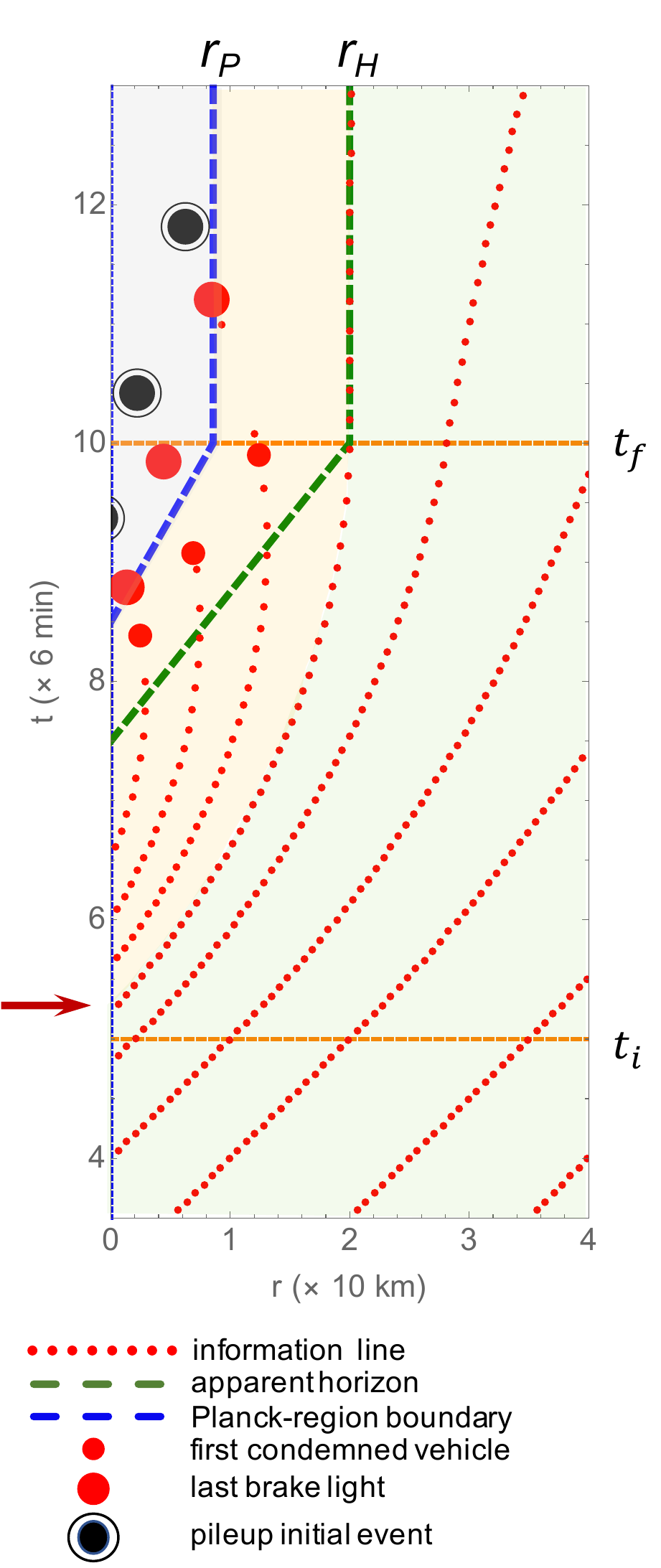}
        \caption{(Color online) Diagram depicting the propagation of the brake-light signal (dotted lines) assuming deterioration of the visual conditions in the period $(t_i,t_f)$. Each dot represents the brake lights of 10 cars. The arrow indicates the rise of the event horizon, which grows up to $r=r_H$. The apparent horizon emerges at $t=(t_i+t_f)/2$ and increases up to merging with the event horizon at $t=t_f$. Inside the apparent horizon, outgoing information lines flow inwards, breeding vehicular-pileup events. The brake lights of the first condemned vehicle are represented by small disks. The exhibited pileups occur in the effective Planck region (or beyond: $r<0$). This plot assumes the parameters of Fig.~\ref{GRAF-3} complemented by $|v_r| = 40~{\rm km/h}$ and $a = 240~{\rm km/h^2}$. }
       \label{GRAF7}
    \end{figure}
    
Let us suppose that the visibility conditions evolve continuously from clear weather at $t<t_i$ to foggy weather at $t>t_f$. Recalling that our information-flow analysis is restricted to the domain $r\geq0$, let us write the reaction time outside the Planck region, where drivers take a finite time to react, as 
    \begin{equation}
       T (t,r)\! = \!\frac{D}{C} \times
    \left\{
	    \begin{array}{lc}
		    \!\!  \left[1 + \frac{|v|}{C}\right]^{-1}  
		    & (t < t_i) \\
		    \\
    	    \!\!  \left[ 1 - \frac{2r_H}{r + r_H} \frac{t - t_i}{ t_f -t_i} + \frac{|v|}{C} \right]^{-1}  
    	    & (t_i\! \leq \! t \! \leq t_f),  \\
    	    \\
	        \!\!  \left[ \frac{r - r_H}{r + r_H} + \frac{|v|}{C} \right]^{-1}  
	        & (t  > t_f)
	    \end{array}
    \right.
    \label{Delay_Trans}
    \end{equation}
while $T(t,r)$ is assumed to be arbitrarily large inside the Planck region. As before, the boundary of the Planck region is defined by the line where $T(t,r)$ diverges (see Fig.~\ref{GRAF7}).  Note that 
$$
T(t,r) = {\rm const}< \infty,\quad {\rm for}\quad t<t_i,
$$ 
while $T(t,r)$ reproduces Eq.~\eqref{delay_E-F} for $t>t_f$. In the region $t_i\leq t\leq t_f$, the $T(t,r)$ function is designed to connect continuously the previous asymptotic regions. 

In Fig.~\ref{GRAF7}, we exhibit the information flow diagram for the nonstationary case. The propagation of the brake-light signals is depicted accordingly by dotted lines.  For $t<t_i$, the information flow is described by homogeneously-distributed straight dotted lines, corresponding to an effective Minkowski spacetime. However, as the weather conditions deteriorate, starting at $t=t_i$, the drivers react slower and an event horizon appears (see arrow), growing up to $r=r_H$ at $t=t_f$. The information propagation lines coincide with the ones shown in  Fig~\ref{GRAF-3} for $t>t_f$. Inside the event horizon, the information lines cannot go far away, in contrast to the ones outside the event horizon. The event horizon contains the apparent horizon, which emerges at $t_{AH} = (t_i + t_f)/2$ and increases until it merges with the event horizon at $t=t_f$. Inside the apparent horizon, outgoing information flows inwards. The boundary of the Planck region is denoted by the internal dashed line which equals $r=r_P$ in the stationary region, $t>t_f$. The spots where the vehicles pile up lie inside the Planck region (or beyond: $r<0$) in Fig.~\ref{GRAF7}.

    %%%%%%%%%%%%%%%%%%%%%%%%%%%%%%%%%%%%%%%%%%%%%%%%%%%%%%%%%%%%%%%%%%%%%
     \section{Summary and final remarks}
     \label{results}
    %%%%%%%%%%%%%%%%%%%%%%%%%%%%%%%%%%%%%%%%%%%%%%%%%%%%%%%%%%%%%%%%%%%%%

In summary, let us assume a convoy of vehicles moving to the left, as in Fig.~\ref{carros2}, and that the first vehicle begins to brake at~$r=0$, before the formation of the event horizon (below the arrow in Fig.~\ref{GRAF7}). In this case, all the following cars have time to react properly, secure the reduced speed $|v_r|$, and pass safely through the foggy region. On the other hand, if the first vehicle applies the brakes after the formation of the event horizon (above the arrow), the future development will be entirely different. The outgoing information may still flow outwards for awhile depending on whether the brakes are applied before the formation of the apparent horizon, but, eventually, will flow inwards as it enters the apparent horizon, leading to accidents. The first condemned car of the convoy is represented by the small disk, which will be rammed by the one behind it. Some other pairwise shocks will occur on the track further but the vehicle responsible for the massive accident will be the one that receives the brake signal from the car ahead (large disk) inside the Planck region. Inside this region, the visibility conditions are so impaired the driver does not react to the brake signal, ramming the car straight ahead (see the target mark in Fig.~\ref{GRAF7}). This is the spot where the rest of the convoy will inadvertently pile up. 

As a result, accidents could be avoided if the drivers were warned to diminish the velocity before entering the event horizon. For this purpose, templates of risky areas should be prepared in advance to forecast where the event horizon would be formed as the visibility conditions deteriorate.

    %%%%%%%%%%%%%%%%%%%%%%%%%%%%%%%%%%%%%%%%%%%%%%%%%%%%%%%%%%%%%%%%
  \begin{acknowledgments}
    L.~S. and G.~M. were fully and partially supported by S\~ao Paulo Research Foundation (FAPESP) under Grant 2019/18616-4 and Conselho Nacional de Desenvolvimento Cient\'\i fico e Tecnol\'ogico (CNPq) under Grant 301544/2018-2, respectively.
  \end{acknowledgments}
  
   \vskip 1 truecm
  {\Large {\bf
  The authors have no conflicts to disclose.}}

%%%%%%%%%%%%%%%%%%%%%%%%%%%%%%%%%%%%%%%%%%%%%%%%%%%%%%%%%%%%%%%%
\end{document}